\documentclass[11pt,twoside]{article}

\usepackage{asp2006}
\usepackage{epsf}
\usepackage{psfig}
\usepackage{lscape}

\begin{document}
\title{Orbits in corotating and counterrotating double bars}
\author{Witold Maciejewski}   
\affil{Astrophysics Research Institute, Liverpool John Moores University, 
Twelve Quays House, Egerton Wharf, Birkenhead CH42 1LD}    
\begin{abstract} 
The backbone of double bars is made out of double-frequency orbits, and loops,
their maps, indicate the bars' extent, morphology and dynamics.
\end{abstract}
Double bars are systems where a nuclear bar is nested inside the main 
bar. The potential of a single bar is constant in its corotating frame, and 
regular closed periodic orbits in that frame can tell us about the extent, morphology, 
and dynamics of the bar. If in a doubly barred galaxy the two bars rotate 
independently, then there is no frame, in which its potential is constant: 
double bars are an example of a periodically varying potential.
Maciejewski \& Sparke (1997, 2000) and Maciejewski \& Athanassoula (2007) 
showed that in such potentials double-frequency orbits (not closed) 
map onto closed curves called loops, which indicate the extent, morphology 
and dynamics of double bars, as closed periodic orbits do in a single bar. 

In a dynamically plausible model of double bar (Maciejewski \& Sparke 2000),
where loops support the whole extent of both bars as they rotate through each 
other, the outer bar extends to its corotation, but the inner bar cannot, 
and it ends at slightly less than half of its corotation radius. 
In this model, loops supporting the inner bar are thicker at
bars parallel, and thinner at bars perpendicular. This 
suggests that the two bars should not rotate through each other
as rigid bodies, but the inner bar should pulsate. Also, loops are
leading the inner bar on its way from parallel with the outer bar to 
perpendicular, while they trail it on its way back to parallel. 
Thus a self-consistent inner bar should not rotate at a constant rate: 
the bars should quickly go through the alignment, and they should spend more 
time being nearly-orthogonal. Recently, all these predictions were fully 
confirmed by N-body models (Debattista \& Shen 2007).

\begin{figure}[t]
\plotone{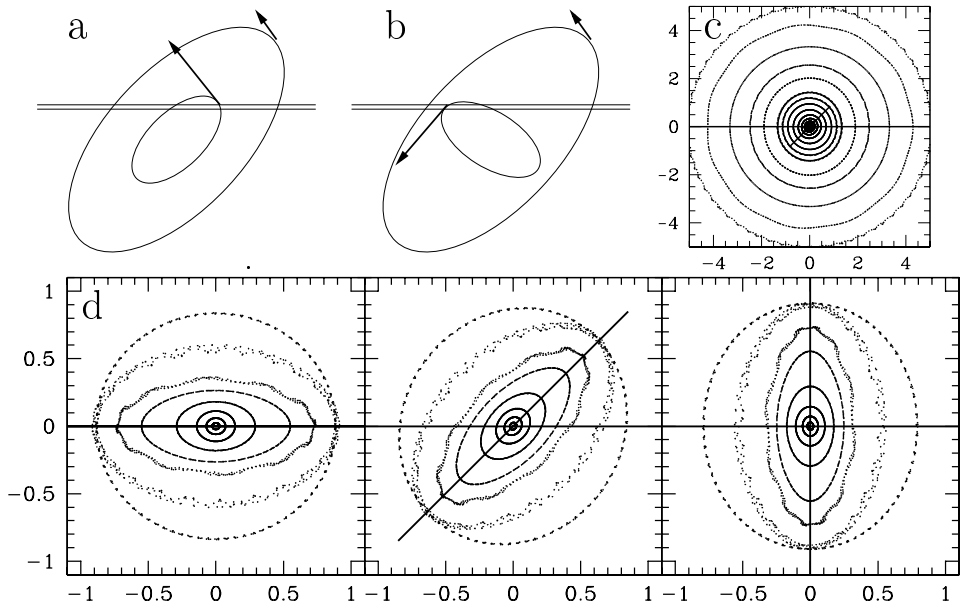}
\caption{\small {\bf a, b:} Two relative positions of the bars (ellipses),
with rotation indicated by arrows. Double line marks the slit. {\bf c:} 
Retrograde loops in two prograde bars. {\bf d:} Retrograde loops in outer 
prograde and inner retrograde bar. Major axes of the bars are drawn, 
the outer bar is always horizontal.}
\end{figure}

In NGC 2950, the Tremaine-Weinberg (TW) integrals (Tremaine \& Weinberg 1984) 
for slits passing through the inner bar fall on a line of higher slope than 
that for the outer slits (Corsini et al.~2003). If the bars were nearly 
parallel (like in Fig.1a), this 
could suggest a faster rotating inner bar, since the contribution from the 
inner bar is likely to boost the integrated velocity (the kinematic integral). 
However, the morphology of NGC 2950 is closer to that from Fig.1b. Then the 
contribution from a fast prograde inner bar should rather bring the kinematic
integral closer to zero. Maciejewski (2006) used the orbital model above to 
show that a fast prograde inner bar cannot reproduce the
observed behaviour of the TW integrals in NGC 2950. Recently, Shen \& 
Debattista (2007) reached the same conclusion based on N-body simulations.

Can the inner bar in NGC 2950 be counterrotating? If there are only particles 
on prograde orbits in the disc, then loops that support a counterrotating bar
should behave like the retrograde x4 orbits in the frame of that bar. Like the 
x4
orbits, retrograde loops in prograde double bars are almost circular (Fig.1c),
hence they cannot support the formation of a retrograde bar in a fully prograde
disc. However, if there is a retrograde population in the disc, then the
retrograde inner bar can be built out of it. In that case, the prograde 
orbits in the outer bar correspond to the retrograde ones in the inner bar
and vice versa. Retrograde loops in double bars, where the outer bar is 
prograde and the inner bar retrograde, support the shape of the inner bar 
throughout its extent, and they rotate with that bar, showing no signs of 
pulsation or acceleration (Fig.1d). Thus counterrotating double bars should
rotate fairly rigidly through each other, keeping the linear relations between 
the TW integrals, observed in NGC 2950, which otherwise may be erased by 
the pulsation of corotating bars (Shen \& Debattista 2007). Since the TW 
integrals utilize the integrated signal, it may be easier to detect 
counterrotation from them than directly in velocity maps.
This work was supported by the Polish KBN grant 1 P03D 007 26.

\end{document}